\documentclass[12pt]{article}
\usepackage{a4wide,amssymb}
\pdfoutput=1
\usepackage{a4wide,amssymb,graphicx}
\usepackage{epsfig}
\usepackage[usenames,dvipsnames]{color}
\usepackage{slashed}
\newcommand{\be}{\begin{equation}}
\newcommand{\ee}{\end{equation}}
\newcommand{\bea}{\begin{eqnarray}}
\newcommand{\eea}{\end{eqnarray}}

\def\circa#1{\,\raise.3ex\hbox{$#1$\kern-.75em\lower1ex\hbox{$\sim$}}\,}

\begin{document}

\begin{titlepage}
%
%


%

\begin{centering}
\vspace{1cm}
{\Large {\bf Resonant SIMP dark matter }} \\

\vspace{1.5cm}

{\bf Soo-Min Choi  and  Hyun Min Lee$^*$}
\vspace{.5cm}

{\it Department of Physics, Chung-Ang University, Seoul 156-756, Korea.} 
\\

\end{centering}
\vspace{2cm}

\begin{abstract}
\noindent
We consider a resonant SIMP dark matter in models with two singlet complex scalar fields charged under a local dark $U(1)_D$. After the $U(1)_D$ is broken down to a $Z_5$ discrete subgroup, the lighter scalar field becomes a SIMP dark matter which has the enhanced $3\rightarrow 2$ annihilation cross section near the resonance of the heavier scalar field. Bounds on the SIMP self-scattering cross section and the relic density can be fulfilled at the same time for perturbative couplings of SIMP.
A small gauge kinetic mixing between the SM hypercharge and dark gauge bosons can be used to make SIMP dark matter in kinetic equilibrium with the SM during freeze-out.

\end{abstract}

\vspace{5cm}

\begin{flushleft}
$^*$Email: hminlee@cau.ac.kr 
\end{flushleft}

\end{titlepage}

\section{Introduction}

Indirect evidences for dark matter are increasing in both diversity and precision, as observed in Cosmic Microwave Background anisotropies and missing masses of galaxies and galaxy clusters, etc.
Thus, dark matter has been one of the driving forces for going beyond the Standard Model (SM), mostly, under the name of the Weakly Interacting Massive Particles (WIMP).
WIMP of weak-scale mass could be a natural outcome of the solution for the hierarchy problem in the SM, but there has been no conclusive hint for WIMP or new particles of weak scale yet in many indirect and direct searches on Earth or in satellites.
On the other hand, light dark matter of sub-GeV scale mass might have been elusive and less explored in previous searches, so it is important to devote more efforts to building consistent scenarios for that possibility and testing them by experiments.  

Strongly Interacting Massive Particles (SIMP) \cite{simp1} have been recently suggested as an alternative thermal dark matter, the relic abundance of which is determined from the freeze-out of the $3\rightarrow 2$ annihilation of dark matter, instead of the $2\rightarrow 2$ annihilation.  The SIMP mechanism is based on the assumption that the $2\rightarrow 2$ annihilation is suppressed and dark matter is in kinetic equilibrium with the thermal plasma at the time of freeze-out \cite{simp1,simp2a,z3dm,simp3}. 
Concrete models for SIMP dark matter have been proposed in the literature \cite{simp2,simp2a,z3dm,globalz3,others} and a review on SIMP dark matter can be found in Ref.~\cite{simp-review}. 
SIMP dark matter typically has a sub-GeV mass and a large self-scattering cross section about $\sigma_{\rm DM}/m_{\rm DM}\sim 1\,{\rm cm^2/g}$, unlike the WIMP case. Then, although such a large self-scattering cross section is constrained by Bullet cluster \cite{bullet} and spherical halo shapes \cite{haloshape}, it can lead to distinct signatures in galaxies and galaxy clusters, such as the off-set of the dark matter subhalo from the galaxy center, as hinted in Abell 3827 \cite{abell3827}.

We briefly review on the production mechanism of SIMP dark matter. 
First, the Boltzmann equation for the SIMP number density $n_{\rm DM}$ includes the additional terms from the $3\rightarrow 2$ annihilation processes as follows,
\bea
\frac{d n_{\rm DM}}{dt}+3H n_{\rm DM}&=&-\langle\sigma v^2\rangle_{3\rightarrow 2} \,(n_{\rm DM}^3-n_{\rm DM}^2 n^{\rm eq}_{\rm DM} )  \nonumber \\
&&-\langle \sigma v\rangle_{2\rightarrow 2}\,(n^2_{\rm DM}- (n^{\rm eq}_{\rm DM})^2)
\eea
where $\langle\sigma v^2\rangle_{3\rightarrow 2}\equiv  \frac{\alpha^3_{\rm eff}}{m^5_{\rm DM}}$ is the effective $3\rightarrow 2$ annihilation cross section and $\langle \sigma v\rangle_{2\rightarrow 2}$ is the $2\rightarrow 2$ annihilation cross section into a pair of SM particles.
When $3\rightarrow 2$ annihilation is dominant, the Boltzmann equation can be rewritten in terms of the DM abundance, $Y=n_{\rm DM}/s$, as
\be
\frac{dY}{dx}=-\lambda \langle \sigma v^2\rangle_{3\rightarrow 2}\, x^{-5} (Y^3-Y^2 Y_{\rm eq})
\ee
with $\lambda\equiv s^2(m_{\rm DM}) /H(m_{\rm DM})$ where $s(m_{\rm DM})=\frac{2\pi^2}{45} g_{*s} m^3_{\rm DM}$ and $H(m_{\rm DM})=\sqrt{\frac{\pi^2}{90} g_*} \,\frac{m^2_{\rm DM}}{M_P}$.  
Then, the approximate solution to the Boltzmann equation leads to the DM relic density as
\bea
\Omega_{\rm DM} h^2&=&\frac{1.05 \times 10^{-10}\,{\rm GeV}^{-2}}{\Big(g^{3/2}_* \frac{m^2_{\rm DM}}{M_P}\int^\infty_{x_F} dx\, x^{-5}\langle \sigma v^2\rangle_{3\rightarrow 2}  \Big)^{1/2}}.
\eea
Therefore, assuming that the $3\rightarrow 2$ annihilation cross section is assumed to be s-wave and taking $H(T_F)=n^2_{\rm DM} \langle \sigma v^2\rangle$ at freeze-out, the SIMP relic density condition is satisfied for $m_{\rm DM}=\alpha_{\rm eff}[0.17g^2_{*s}/(x^4_F g^{1/2}_*) T^2_{\rm eq} M_P]^{1/3}$ where $T_{\rm eq}$ is the temperature at matter-radiation equality given by $T_{\rm eq}=0.8\,{\rm eV}$ and $g_*, g_{*s}$ are the effective numbers of relativistic species in radiation and entropy densities, respectively.
Then, for $x_F\equiv \frac{m_{\rm DM}}{T_F}\approx 20$ and $g_*=10.75$,  we get $m_{\rm DM}\approx( 35\,\alpha_{\rm eff}) \,{\rm MeV}$.
Thus, we need to choose $\alpha_{\rm eff}=1-30$ for the SIMP mass being in the range between $35\,{\rm MeV}$ and $900\,{\rm MeV}$. Consequently, the fact that the correct relic density for SIMP requires such a large effective DM coupling  could be in a tension with the validity of perturbativity and unitarity \cite{globalz3,z3dm,sannino}. Furthermore, since the DM self-scattering cross section behaves as $\sigma_{\rm self}\sim \frac{\alpha^2_{\rm eff}}{m^2_{\rm DM}}$, the resultant large effective DM coupling is also constrained by the Bullet cluster or spherical halo shapes.

In this article, we consider a novel possibility to generate the tree-level 5-point interaction for dark matter by exchanges of a heavy field in a local dark $U(1)_D$ model with two complex singlet scalar fields in the dark sector. The resonant enhancement of the $3\rightarrow 2$ annihilation can tolerate the necessity of introducting large couplings and avoid the strong constraints from the Bullet cluster and halo shapes as well as unitarity bounds.  We show how the parameter space of scalar interactions is constrained by perturbativity/unitarity and bounds on self-interactions and discuss how the resonant SIMP dark matter can be searched for.

\section{A gauged $Z_5$ symmetry}

We introduce a $U(1)_D$ gauge symmetry which is broken down to a $Z_5$ discrete subgroup due to the  VEV of a complex singlet scalar $\phi$ carrying a charge $q_\phi=+5$ under $U(1)_D$. On the other hand, assumed that a complex scalar $\chi$ carries a charge $q_\chi=+1$ under $U(1)_D$ and it does not get a VEV,  it can be a candidate for stable dark matter. In order for $\chi$ to be a SIMP dark matter, we need to induce a 5-point interaction for $\chi$ but there is no such interaction at tree level due to the remaining $Z_5$ symmery. 
Therefore, we introduce an additional singlet scalar $S$ carrying a charge $q_S=+3$ under $U(1)_D$.
$U(1)_D$ charges are given in Table~1.
Then, after integrating out the scalar field $S$, we can obtain an effective 5-point self-interaction, $\chi^5$, respecting a $Z_5$ discrete symmetry.  If the additional scalar field $S$ is not decoupled, there is a possibility that the resulting $3\rightarrow 2$ annihilation for dark matter can be enhanced due to the resonance of the scalar field $S$. Moreover, the scalar field $S$, if lighter than $\chi$, can be a SIMP dark matter too and its $3\rightarrow 2$ annihilation can be enhanced in a similar matter at the resonance of the heavier scalar field $\chi$.  

\begin{table}[ht]
\centering
\begin{tabular}{|c||c|c|c|}
\hline 
& $\phi$ &  $S$  & $\chi$   \\ [0.5ex]
\hline 
$U(1)_D$ & $+5$ & $+3$ & $+1$ 
 \\ [0.5ex]
\hline
\end{tabular}
\caption{$U(1)_D$ charges.}
\label{table:charges1}
\end{table}

The Lagrangian for two singlet complex scalars, $\chi$ and $S$, dark Higgs $\phi$ and dark gauge boson $V_\mu$, in our model, is given by
\bea
{\cal L}_{\rm hid}=-\frac{1}{4}V_{\mu\nu} V^{\mu\nu}+ |D_\mu\phi|^2 + |D_\mu \chi|^2 +|D_\mu S|^2 - V_{\rm hid} \nonumber 
\eea
where the field strength tensor for dark photon is $V_{\mu\nu}=\partial_\mu V_\nu-\partial_\nu V_\mu$, and covariant derivatives are $D_\mu \phi=(\partial_\mu - iq_\phi g_D V_\mu)\phi$,  
$D_\mu\chi = (\partial_\mu - i q_\chi g_D V_\mu)\chi$, and $D_\mu S=(\partial_\mu-i q_S g_D V_\mu)S$ with $q_\phi=+5, q_\chi=+1$, $q_S=+3$ and $g_D$ being dark gauge coupling. The scalar potential in the hidden sector is $V_{\rm hid}$ is given by 
 \bea
V_{\rm hid}&=& -m^2_\phi |\phi|^2 + m^2_{\chi} |\chi|^2+m^2_S |S|^2 \nonumber \\
&& +\lambda_\phi |\phi|^4 +\lambda_\chi |\chi|^4 +\lambda_S |S|^4  \nonumber \\
&&+ \lambda_{\phi \chi} |\phi|^2 |\chi|^2+ \lambda_{S \chi}|S|^2 |\chi|^2 + \lambda_{\phi S}|\phi|^2 |S|^2  \\
&&+\frac{1}{\sqrt{2}}\lambda_1 \phi^\dagger S^2 \chi^\dagger+ \frac{1}{\sqrt{2}}\lambda_2 \phi^\dagger S\chi^2  +\frac{1}{6}\lambda_3 S^\dagger\chi^3+{\rm h.c.}.   \nonumber \label{dmcouplings} 
\eea
We note that there can be extra quartic couplings between the SM Higgs doublet $H$ and the singlet scalars, e.g. $|H|^2|\chi|^2$, but we assume that they are suppressed enough to satisfy the bounds on the invisible decay of Higgs boson \cite{Hinvisible,z3dm}. Thus, we don't consider extra quartic couplings any more in the following discussion. 

After expanding the dark Higgs $\phi$ around a nonzero VEV as $\phi=\frac{1}{\sqrt{2}}(v_D+h_D)$, the renormalizable interactions between $\chi$ and $S$ in eq.~(\ref{dmcouplings}) become
\bea
{\cal L}_{S,\chi}=-\frac{1}{2}\lambda_1 v' S^2 \chi^\dagger -\frac{1}{2}\lambda_2 v' S\chi^2-\frac{1}{6}\lambda_3 S^\dagger \chi^3 +{\rm h.c.}.  \label{dmcouple}
\eea
Therefore,  the resulting cubic and quartic couplings respect the $Z_5$ discrete symmetry and are responsible for generating the 5-point interactions for $\chi$ or $S$. Moreover, $\lambda_{1,2}$ in the cubic interactions are responsible for the decay of the heavier scalar to a pair of the lighter ones, if kinematically allowed. 
On the other hand, the dark gauge boson gets mass, $m_V=5 g_D v_D$, due to the $U(1)_D$ breaking, and it can couple to the charged particles in the SM through the gauge kinetic mixing and thus play a role of messenger between dark matter and the SM.

\section{Resonant enhancement of the $3\rightarrow 2$ annihilation}

We assume that the dark Higgs and the dark gauge boson are heavier than dark matter such that their contributions to the annihilation of dark matter are suppressed.
However, $Z'$ gauge boson contributes dominantly to the kinetic scattering between SIMP dark matter and the SM charged leptons \cite{simp2a,z3dm,simp3}.

\begin{figure*}[!t]
  \begin{center}
   \begin{tabular}{cccccccc} 
   \includegraphics[height=0.12\textwidth]{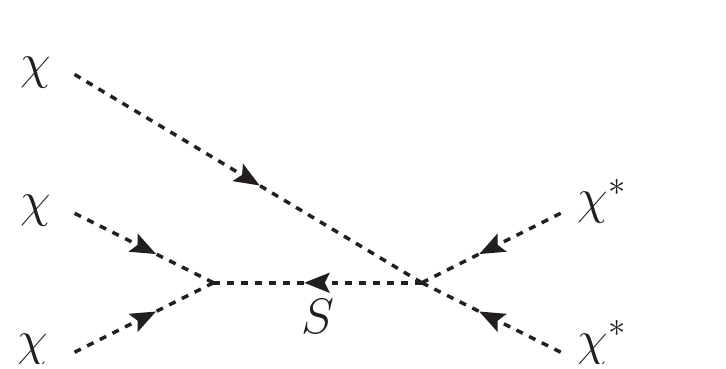}
   &
   \includegraphics[height=0.12\textwidth]{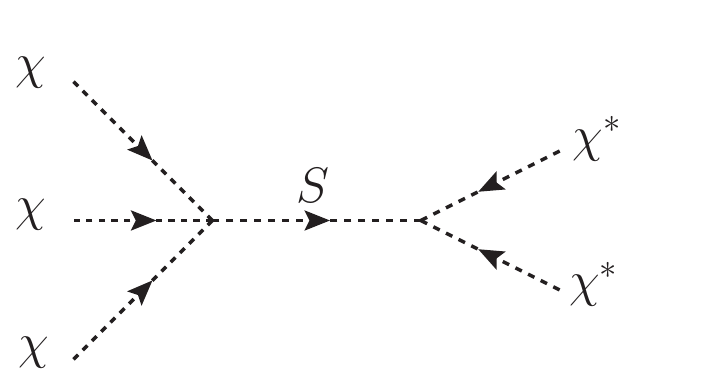}
 &
 \includegraphics[height=0.12\textwidth]{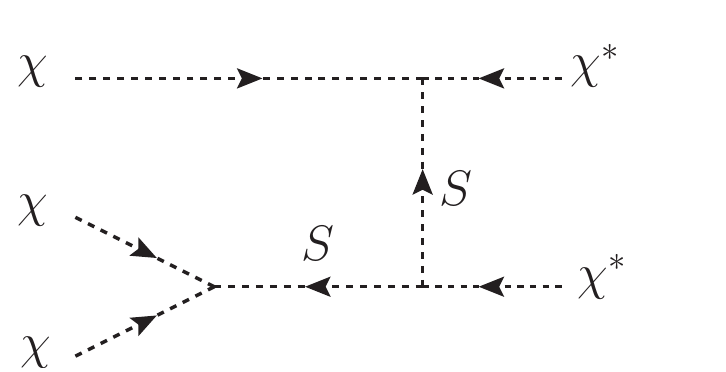}
   &
   \includegraphics[height=0.12\textwidth]{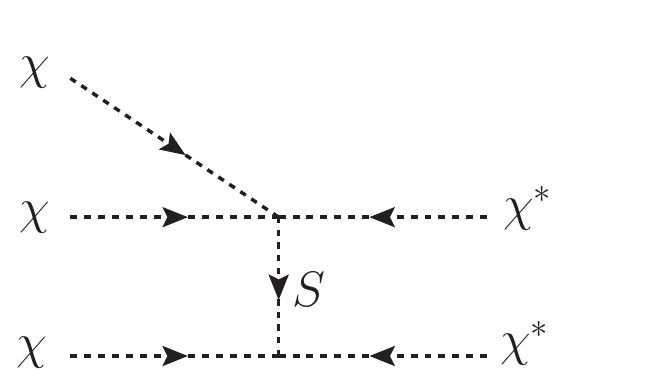}
 \\
 \includegraphics[height=0.12\textwidth]{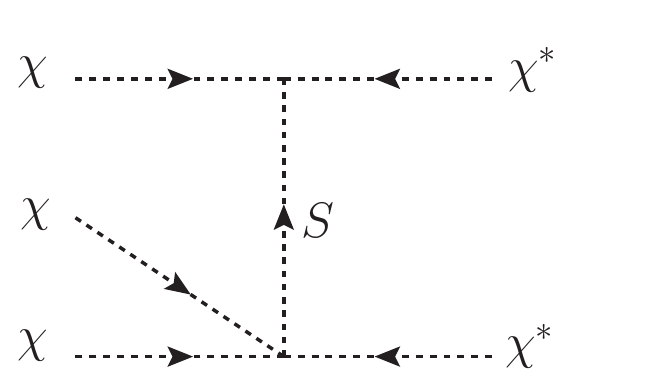}
   &
   \includegraphics[height=0.12\textwidth]{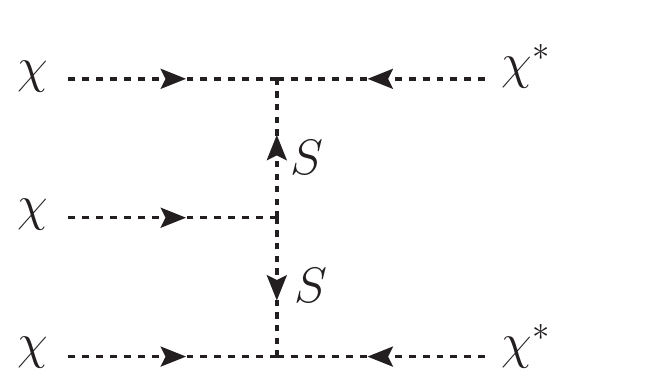}
    &
    \includegraphics[height=0.12\textwidth]{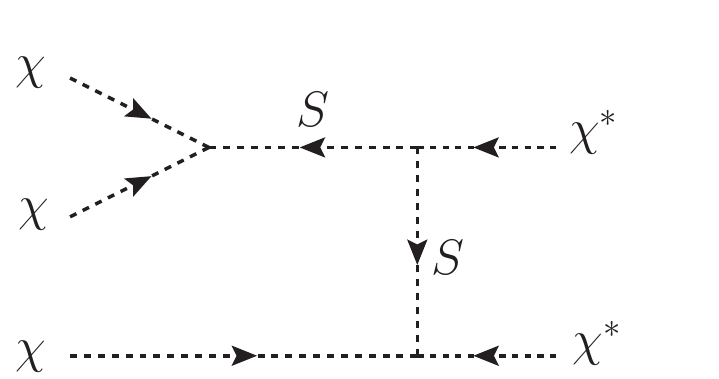}
 &
 \includegraphics[height=0.12\textwidth]{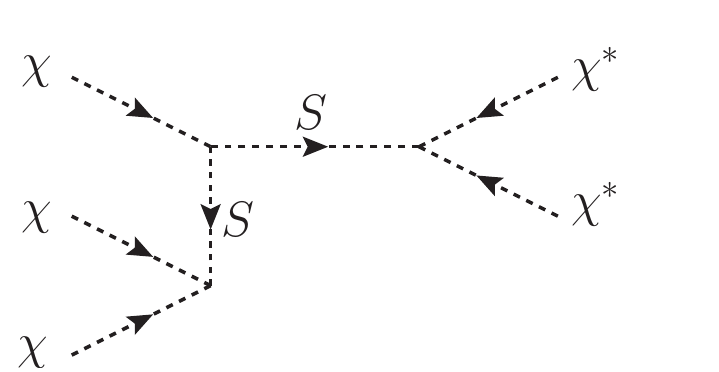}
  \end{tabular}
   \end{center}
  \caption{Feynman diagrams for  $\chi\chi\chi\rightarrow \chi^*\chi^*$. }
  \label{3to2a}
\end{figure*}

First, taking  $m_S> 2m_\chi$, the singlet scalar $S$ decays into a pair of dark matter $\chi$.   
In this case, while the $2\rightarrow 2$ (semi-) annihilation processes in hidden sector are kinematically forbidden, the  $3\rightarrow 2$ process,  $\chi\chi\chi\rightarrow \chi^*\chi^*$,  is a dominant annihilation process. But, the dark Higgs or the dark gauge boson does not contribute to the processes even through intermediate states, unlike the $Z_3$ case \cite{z3dm}.  Moreover, the $3\rightarrow 2$ process for $\chi$ is made possible due to the exchanges of the scalar $S$ as shown in Fig.~\ref{3to2a}.

In the non-relativistic limit for dark matter, the squared amplitude for the $\chi\chi\chi\rightarrow \chi^*\chi^*$ process is
\bea
|{\cal M}_{\chi\chi\chi\rightarrow \chi^*\chi^*}|^2&=&\frac{25m^2_\chi R^2_2}{3}
\bigg| \frac{\lambda_3 (37m^4_\chi-21m^2_\chi m^2_S+2m^4_S)}{(m^2_\chi+m^2_S)(4m^2_\chi-m^2_S+i\Gamma_S m_S)(9m^2_\chi-m^2_S+i\Gamma_S m_S)}  \nonumber \\
&&\quad-\frac{6 m^2_\chi R_1R_2(11m^4_\chi-8m^2_\chi m^2_S+m^4_S) }{(m^2_\chi+m^2_S)^2(4m^2_\chi-m^2_S+i\Gamma_S m_S)(9m^2_\chi-m^2_S+i\Gamma_S m_S)} \bigg|^2  \label{3to2}
\eea
where $R_{1,2}\equiv \lambda_{1,2} v'/(\sqrt{2} m_\chi)$ and  the decay width for $S$ is given by
\be
\Gamma_S=\frac{m^2_\chi R^2_2}{16\pi m_S}\,\Big(1-\frac{4m^2_\chi}{m^2_S}\Big)^{1/2}.
\ee
Then, for CP conservation, the DM number density is given by $n_{\rm DM}=n_\chi+n_{\chi^*}$, for $n_{\chi}=n_{\chi^*}$, and  the effective $3\rightarrow 2$ annihilation cross section is obtained as
\be
\langle\sigma v^2\rangle_{\chi, 3\rightarrow 2}=\frac{\sqrt{5}}{1536\pi m^3_\chi}\,|{\cal M_{\chi\chi\chi\rightarrow \chi^*\chi^*}}|^2 \equiv \frac{\alpha^3_{\rm eff}}{m^5_\chi}.  \label{3to2chi}
\ee 
We note that all the $Z_5$-invariant quartic couplings between $\chi$ and $S$ participate in the $\chi\chi\chi\rightarrow \chi^*\chi^*$ process.

Assuming that the dark Higgs and dark gauge boson are heavy enough and ignoring the mixing quartic coupling between $\chi$ and dark Higgs field, we also obtain the $2\rightarrow 2$  self-scattering processes for $\chi$ as 
\be
\sigma_{\chi,\rm self}= \frac{1}{64\pi m^2_\chi} (|{\cal M}_{\chi\chi}|^2 + |{\cal M}_{\chi\chi^*}|^2). \label{chiDMself}
\ee
with 
\bea
|{\cal M}_{\chi\chi}|^2&=&2\left|2\left(\lambda_\chi+\frac{2g^2_D m^2_\chi}{m^2_V}\right)+\frac{m^2_\chi R^2_2 }{4m^2_\chi-m^2_S+i\Gamma_S m_S} \right|^2, \nonumber  \\
|{\cal M}_{\chi\chi^*}|^2&=&4\left|2\left(\lambda_\chi-\frac{g^2_D m^2_\chi}{m^2_V}\right)-\frac{m^2_\chi R^2_2 }{m^2_S}\right|^2. \nonumber   
\eea
Here, we note that unitarity bounds on self-scattering are $|{\cal M}_{\chi\chi}|, |{\cal M}_{\chi\chi^*}|<8\pi$.

Remarkably, the annihilation cross section for $\chi\chi\chi\rightarrow \chi^*\chi^*$ is then enhanced  
near the resonance with  $m_S= 3m_\chi$, as can be seen in eq.(\ref{3to2chi}) with (\ref{3to2}).  Including a nonzero velocity of dark matter in the center of mass energy $s=9m^2_\chi(1+v^2_{\rm rel}/4)$ in the propagators in eq.~(\ref{3to2}), the annihilation cross section for $\chi\chi\chi\rightarrow \chi^*\chi^*$ before thermal average has a temperature-dependent pole as follows,
\bea
(\sigma v^2)_{\chi} \equiv\frac{c_\chi}{m^5_\chi}\, \frac{\gamma^2_S}{(\epsilon_S-v^2_{\rm rel}/4)^2+\gamma^2_S}
\eea
where $\gamma_S\equiv m_S\Gamma_S/(9m^2_\chi)$,  $\epsilon_S\equiv (m^2_S-9m^2_\chi)/(9m^2_\chi)$ parametrizes the off-set from the resonance pole, and $c_\chi$ is a constant parameter. 
Then, following the similar steps as for WIMP in Ref.~\cite{griest,gondolo,murayama}, we obtain the general result for the thermal-averaged SIMP annihilation cross section as follows,
\bea
\langle(\sigma v^2)_{\chi} \rangle &=& \frac{x^{3/2}}{2\sqrt{\pi}}\int^\infty_0 dv\, v^2 (\sigma v^2)_\chi \, e^{-xv^2/4}  \nonumber \\
&=& \frac{2c_\chi}{m^5_\chi}\, x^{3/2} \sqrt{\pi} \gamma_S \,{\rm Re}\left(z^{1/2}_S  e^{-x z_S} {\rm Erfc}(-ix^{1/2} z^{1/2}_S) \right)
\eea
where $x\equiv m_\chi/T$, $z_S\equiv \epsilon_S+i \gamma_R$ and 
\be
{\rm Erfc}(x)\equiv \frac{2}{\sqrt{\pi}} \int^\infty_x \, e^{-t^2} \,dt. 
 \ee
 In particular, for $\gamma_S\ll  1$, by using the formula,
 \be
\left( \frac{\gamma_S}{x^2+\gamma^2_S}\right)\bigg|_{\gamma_S\ll 1}=\pi \delta(x),
 \ee
 we get the approximate form for the thermal-averaged SIMP annihilation cross section,
 \be
 \langle(\sigma v^2)_{\chi} \rangle\approx \frac{2c_\chi}{m^5_\chi}\, x^{3/2}\sqrt{\pi} \gamma_S \epsilon^{1/2}_S \,e^{-x\epsilon_S}\,\theta(\epsilon_S)
\ee
where $\theta(x)$ is the Heaviside step function with $\theta(x)=1,\, x\geq 0$, and $\theta(x)=0,\,\, x<0$. 
Thus, for $\epsilon_S>0$, i.e. $m_S>3m_\chi$,  the tail of the Maxwell-Boltzmann distribution at large velocities allows for a resonant enhancement of the annihilation cross section.  On the other hand, for $\epsilon_S<0$,  i.e. $m_S<3m_\chi$, the annihilation cross section almost vanishes, because the center of mass with nonzero velocity is always above the resonance. 
 
Consequently, the SIMP relic density can be determined by 
\be
\Omega_\chi  = \frac{m_\chi s_0 /\rho^0_c}{\Big( 2\lambda J(x_f)\Big)^{1/2}}
\ee
where $s_0$ and $\rho^0_c$ are the entropy and critical densities at present, $\lambda\equiv s(m_\chi)^2/H(m_\chi)$ and
\be
J(x_F)\equiv \int^\infty_{x_F} \frac{\langle (\sigma v^2)_\chi\rangle}{x^5}\, dx.
\ee
with $x_F=m_\chi/T_F\simeq 10-20$ at freeze-out temperature.
For $\gamma_S\ll 1$, the $J$ factor is approximated as
\be
J(x_F)\approx \frac{2c_\chi}{m^5_\chi}\, \sqrt{\pi} \gamma_S \epsilon^{1/2}_S \theta(\epsilon_S) F(\epsilon_S). 
\ee
with
\bea
F(\epsilon_S)&\equiv& \int^\infty_{x_F} dx\, x^{-7/2}\, e^{-x\epsilon_S}  \\
&=& \frac{1}{60} \bigg(-32\sqrt{\pi}\epsilon^{5/2}_S \,{\rm Erfc}\Big(x^{1/2}_F \sqrt{\epsilon_S}\Big)+x^{-5/2}_F \, e^{-x_F \epsilon_S} \Big(24+ 8x_F \epsilon_S(-2 + 4x_F \epsilon_S) \Big)  \bigg).  \nonumber
\eea
For instance, $\epsilon_S\lesssim x^{-1}_F$,  we get $F(\epsilon_S)\approx \frac{2}{5} x^{-5/2}_F e^{-x_F \epsilon_S} $. Then, we get $\Omega_\chi/\Omega^0_\chi=\sqrt{J_0/J}\sim (\gamma_S/\epsilon_S)^{1/2}(x_F\epsilon_S)^{-3/2}$ where $\Omega^0_\chi$ and $J_0$ are computed for $v_{\rm rel}=0$ and $\gamma_S\ll \epsilon_S$ is assumed. Therefore, as the resonant enhancement is improved with nonzero temperature taken into account, a smaller SIMP coupling is favored for a correct relic density, being consistent with perturbativity for SIMP dark matter.

\begin{figure*}[!t]
  \begin{center}
   \begin{tabular}{cccccccc} 
   \includegraphics[height=0.43\textwidth]{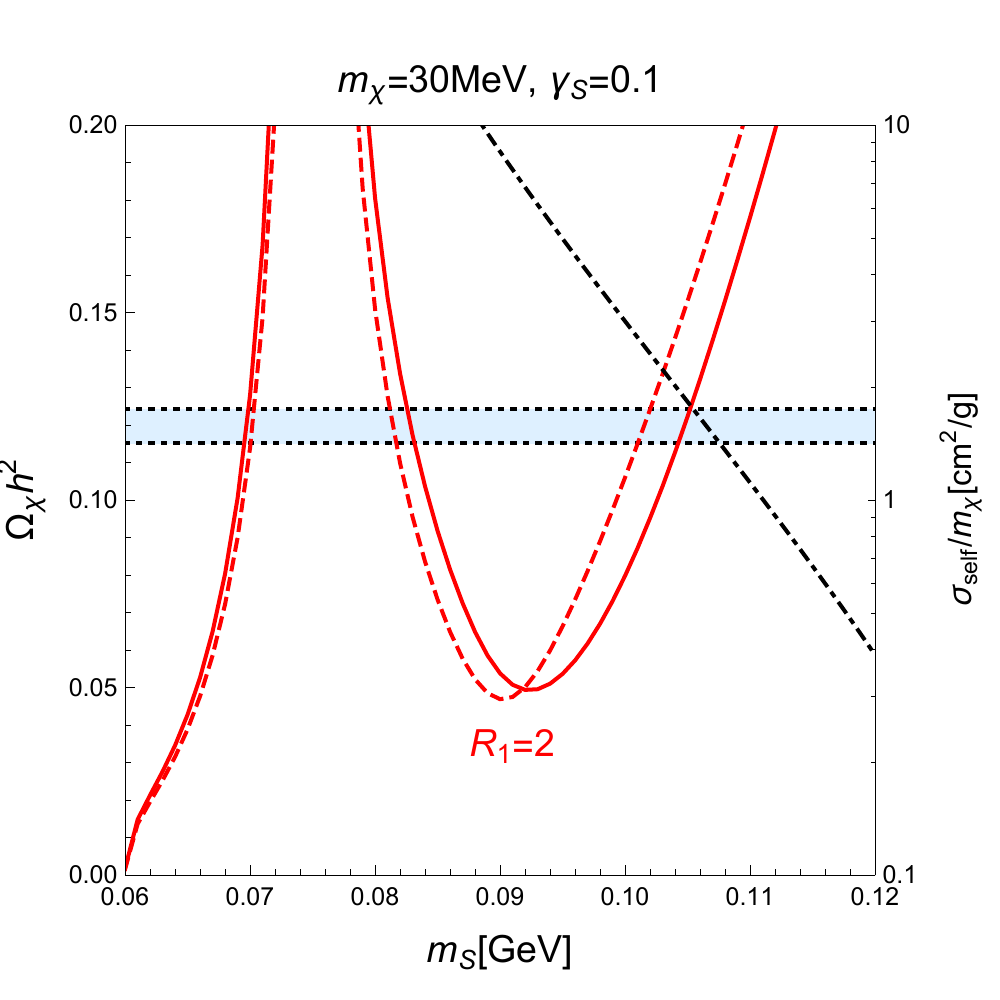}
   \includegraphics[height=0.43\textwidth]{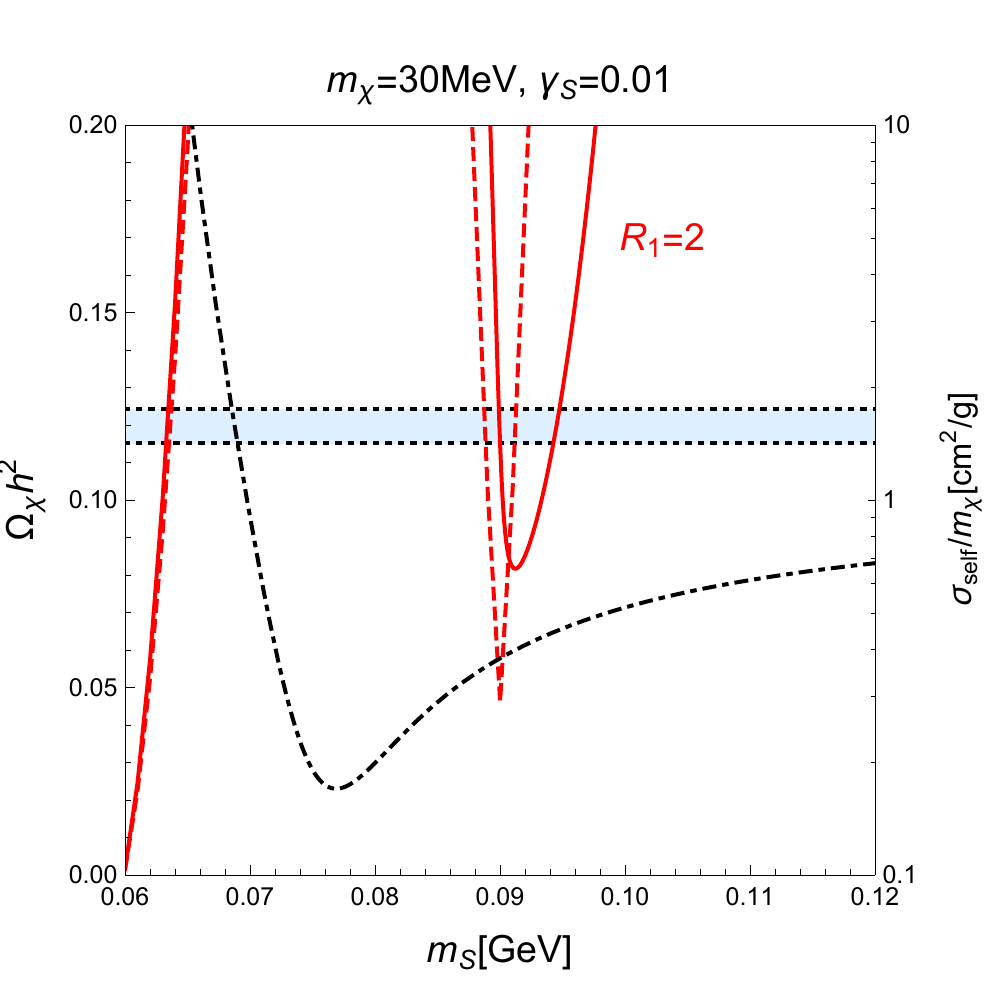}
     \end{tabular}
   \end{center}
  \caption{Temperature effect near the resonance. The relic density of $\chi$ SIMP is given as a function of $m_S$ for zero or nonzero temperature of dark matter in dashed or solid lines. We took $R_1=2$. Planck $3\sigma$ band on the relic density is imposed in horizontal light-blue region.  For $m_V=500\,{\rm MeV}$,  $g_D=0.1$ and $\lambda_\chi=1$, self-scattering cross section ($\sigma_{\rm self}/m_\chi$) is shown in units of $1{\rm cm^2/g}$  in dot-dashed line.
  }
  \label{tempcompare}
\end{figure*}

\begin{figure*}[!t]
  \begin{center}
   \begin{tabular}{cccccccc} 
   \includegraphics[height=0.43\textwidth]{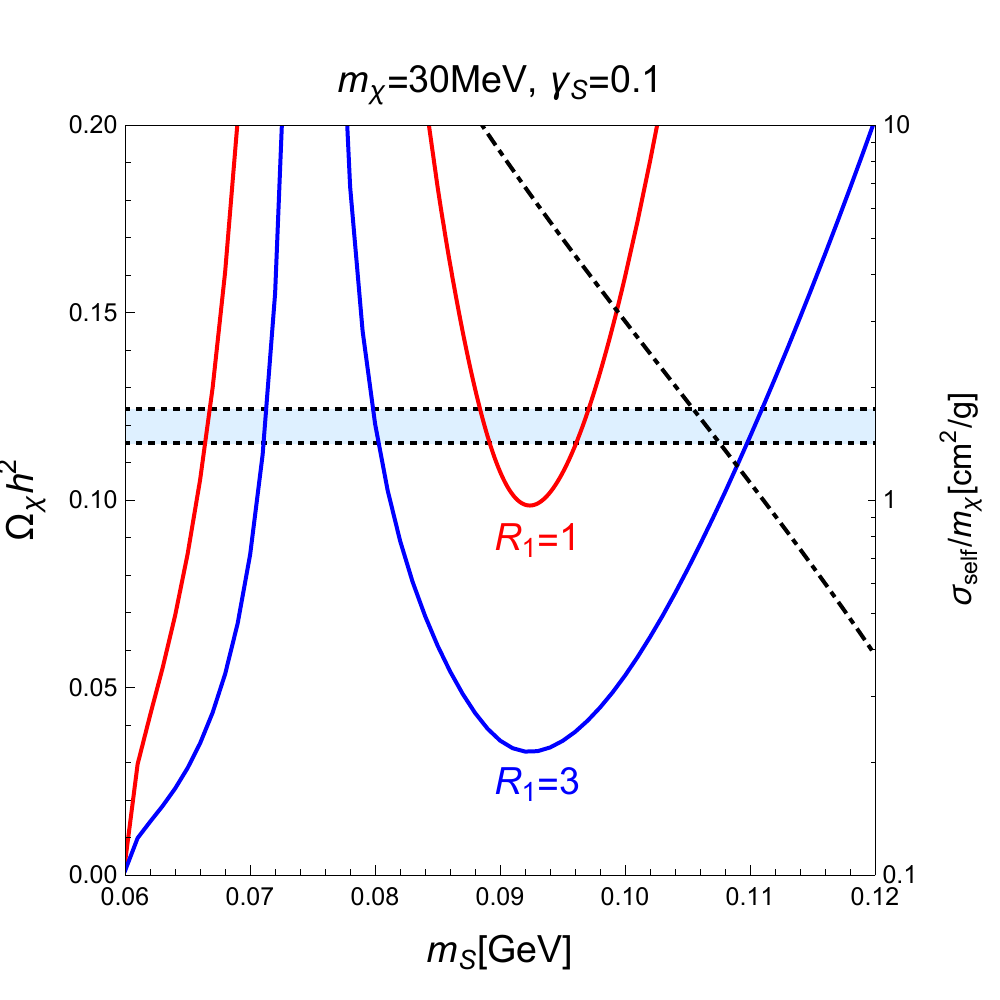}
  \includegraphics[height=0.43\textwidth]{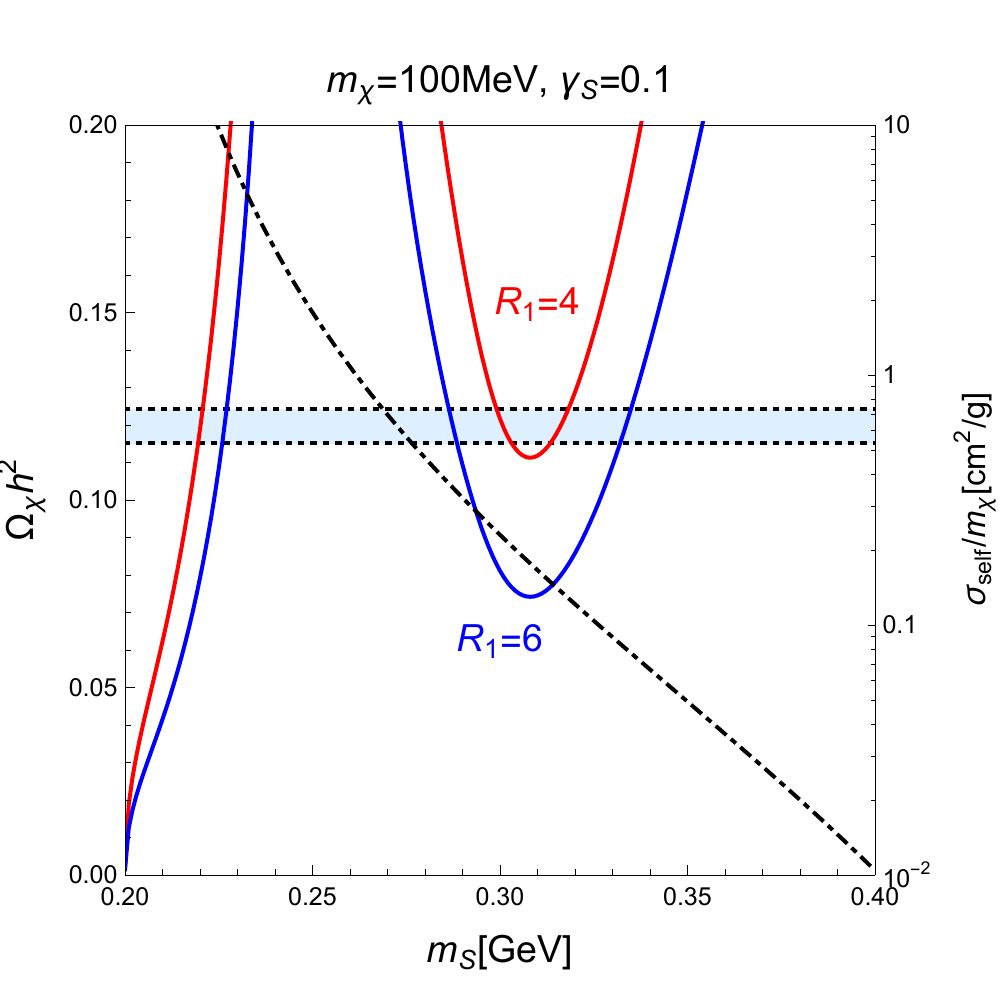} \\
   \includegraphics[height=0.43\textwidth]{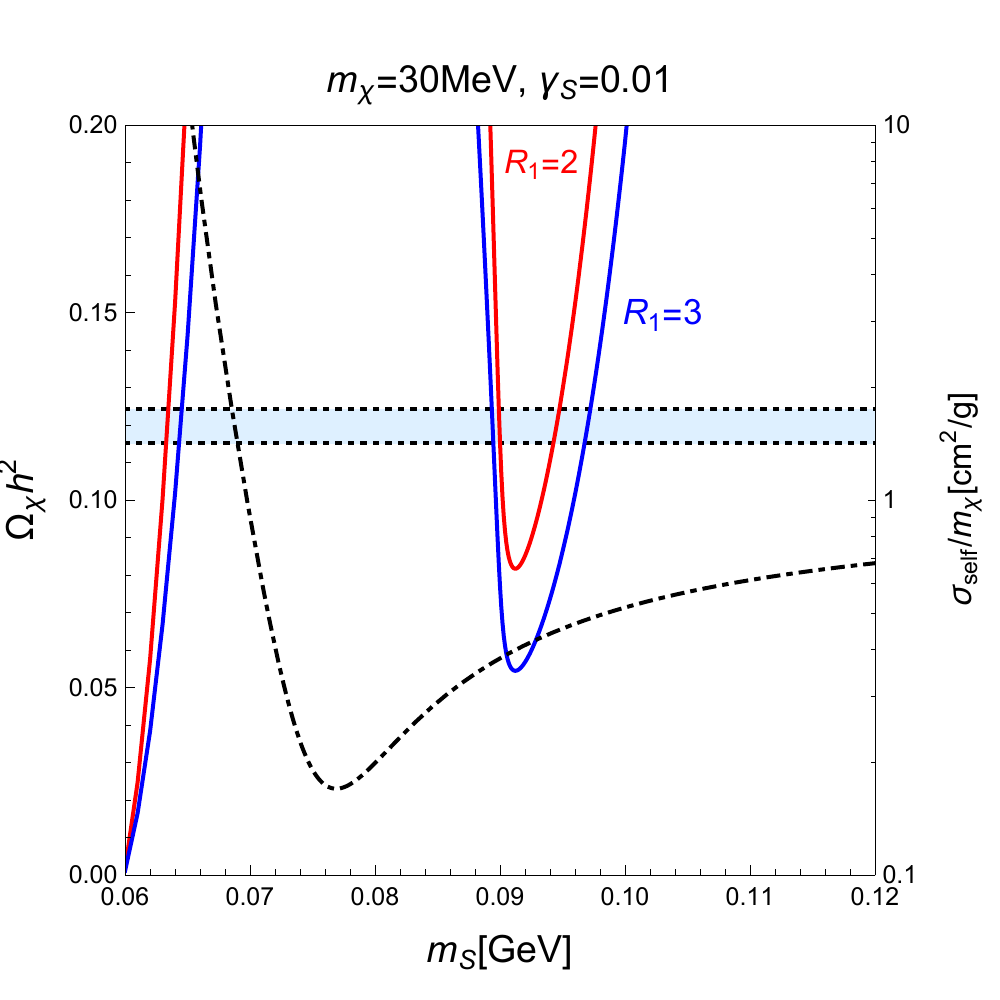}
   \includegraphics[height=0.43\textwidth]{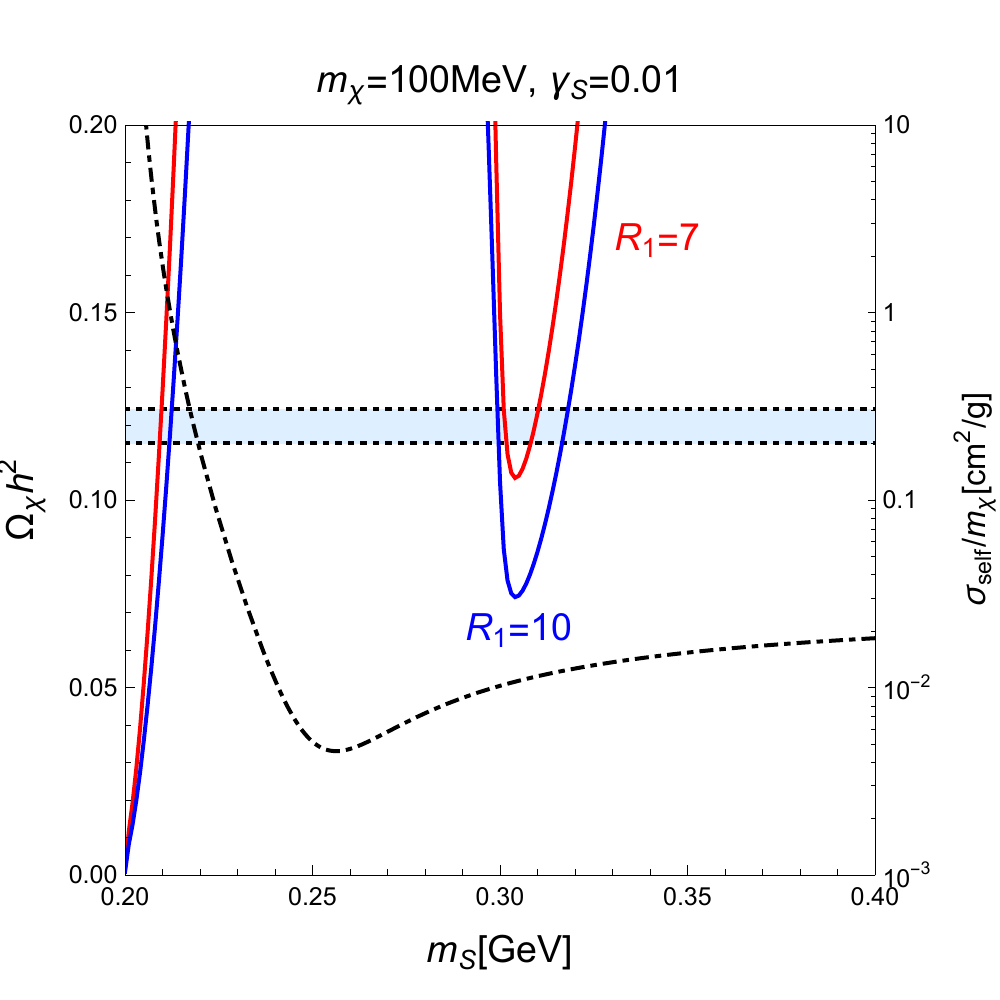}
     \end{tabular}
   \end{center}
  \caption{Relic density of $\chi$ SIMP as a function of the mediator mass $m_S$. We took $R_1$ of order one in solid lines. Planck $3\sigma$ band on the relic density is imposed in horizontal light-blue region.  For $m_V=500\,{\rm MeV}$,  $g_D=0.1$ and $\lambda_\chi=1$, self-scattering cross section ($\sigma_{\rm eff}/m_\chi$) is shown in units of $1{\rm cm^2/g}$  in dot-dashed line.  
    }
  \label{tempchi}
\end{figure*}

For simplicity, we choose $\lambda_3=0$ and compare the relic density of the $\chi$ SIMP dark matter as a function of the resonance mass $m_S$ in Fig.~\ref{tempcompare}. Here, the temperature of dark matter is taken to zero or nonzero near the resonance in dashed or solid lines. 
For nonzero $\lambda_3$, the $3\rightarrow 2$ annihilation cross section gets smaller or larger, depending on whether the sign of $\lambda_3$ is the same as $\lambda_1\lambda_2$ or not, but the results are not different qualitatively.  
In Fig.~\ref{tempchi}, we also show the relic density of the $\chi$ SIMP dark matter, depending on the  SIMP mass and the SIMP coupling, $R_1$ of order one, in solid lines. Consequently, we find that the relic density changes significantly, depending on the mass and width of the resonance\footnote{See Ref.~\cite{griest,gondolo,murayama} for the temperature effect on the resonant enhancement of the relic density for WIMP dark matter. More general discussion on the temperature effects on SIMP dark matter will be published elsewhere \cite{simp-temp}.}.

In both Fig.~\ref{tempcompare} and Fig.~\ref{tempchi} , we depict the self-scattering cross section per SIMP mass, $\sigma_{\rm self}/m_\chi$, in dot-dashed lines, and impose the Planck $3\sigma$ values \cite{planck} on the relic density in light-blue region.  
As a result, the correct relic density can be obtained for the SIMP coupling of order one near the resonance, while the bounds on the self-scattering cross section, $\sigma_{\rm self}/m_\chi<1\,{\rm cm^2/g}$, obtained from Bullet cluster \cite{bullet} and halo shapes \cite{haloshape}, as well as unitartity and perturbativity bounds, are satisfied.  
The smaller the width of the scalar field $S$, the smaller the self-scattering cross section and the larger SIMP masses are allowed to satisfy the correct relic density and the bounds on self-scattering.  
We note that another resonance at $m_S=2m_\chi$ appears in both $2\rightarrow 2$ and $3\rightarrow 2$ processes, so the region near those additional resonances is disfavored by unitarity or bounds on self-scattering.

\begin{figure*}[!t]
  \begin{center}
  \begin{tabular}{cccc} 
\includegraphics[height=0.13\textwidth]{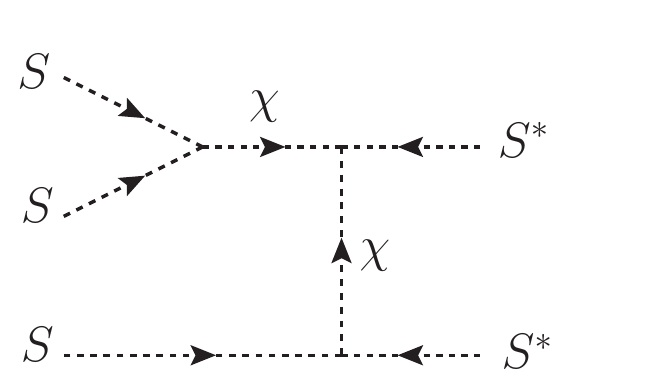}
&
\includegraphics[height=0.13\textwidth]{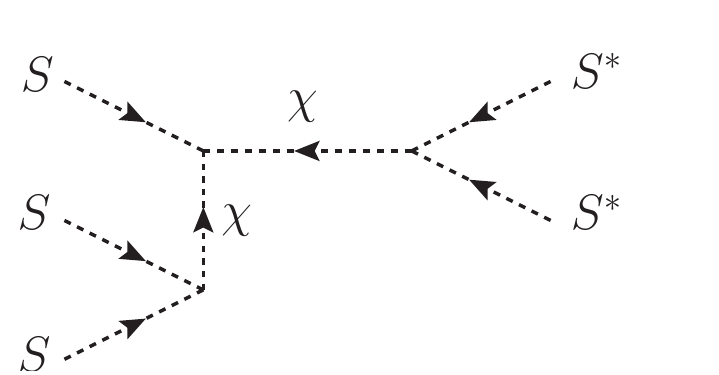}
&
\includegraphics[height=0.13\textwidth]{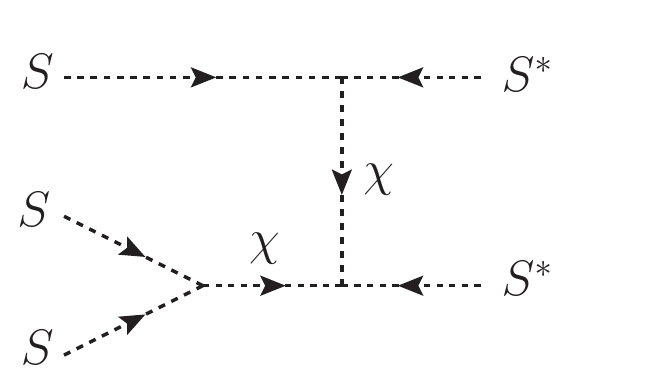}
&
\includegraphics[height=0.13\textwidth]{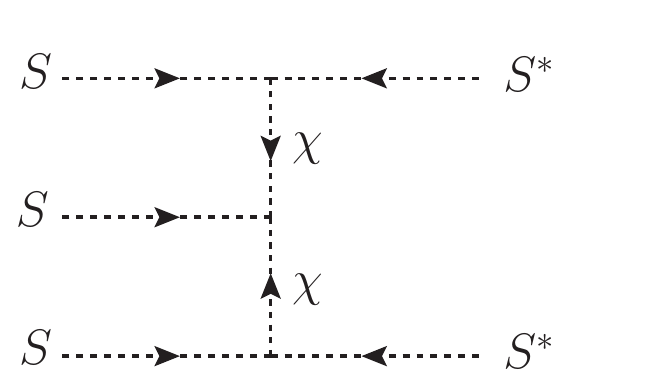}
 \end{tabular}
     \end{center}
  \caption{Feynman diagrams for  $SSS\rightarrow S^* S^*$. }
  \label{3to2add}
\end{figure*}

Now we consider the case with $2 m_S< m_\chi$ for which $\chi$ decays into a pair of $S$ and $S$ is stable.
While the semi-annnihilation channels for $S$ are closed kinematically, the $3\rightarrow 2$ annihilation channel, $SSS\rightarrow S^*S^*$ is possible as shown in Fig.~\ref{3to2add}. 
In the non-relativistic limit for dark matter, the squared amplitude for the $SSS\rightarrow S^*S^*$ process is
\bea
|{\cal M}_{SSS\rightarrow S^*S^*}|^2=\frac{300  R^4_1 R^2_2 (m_\chi^4-8m_\chi^2 m_S^2+11 m^4_S)^2m^6_\chi}{(m_\chi^2+m_S^2)^4[(9m^2_S-m^2_\chi)^2+\Gamma^2_\chi m^2_\chi][(4m^2_S-m^2_\chi)^2+\Gamma^2_\chi m^2_\chi]} \label{3to2s}
\eea
where the decay width for $\chi$ is given by
\be
\Gamma_\chi= \frac{m_\chi R^2_1}{16\pi} \,\Big(1-\frac{4m^2_S}{m^2_\chi}\Big)^{1/2}.
\ee
We note that the above squared amplitude is the same form as eq.~(\ref{3to2} with $\lambda_3=0$ in the case of $\chi\chi\chi\rightarrow \chi^*\chi^*$ process, but with $R_1$ and $m_\chi$ being interchanged by $R_2$ and $m_S$, respectively.
Then, for CP conservation, the DM number density is given by $n_{\rm DM}=n_S+n_{S^*}$, for $n_{S}=n_{S^*}$, and the effective $3\rightarrow 2$ annihilation cross section is obtained as
\be
\langle\sigma v^2\rangle_{S,3\rightarrow 2}=\frac{\sqrt{5}}{1536\pi m^3_S}\,|{\cal M_{SSS\rightarrow S^*S^*}}|^2 \equiv \frac{\alpha^3_{\rm eff}}{m^2_S}.
\ee 
We note that only $\lambda_{1,2}$ quartic couplings between $\chi$ and $S$ participate in the $SSS\rightarrow S^*S^*$ process.

Similarly to the $\chi$ SIMP case, when dark Higgs and dark gauge boson are heavy enough and ignoring the mixing quartic coupling between $S$ and dark Higgs field, the $2\rightarrow 2$  self-scattering processes for $S$ is given by
\be
\sigma_{S,{\rm self}}= \frac{1}{64\pi m^2_S} (|{\cal M}_{SS}|^2 + |{\cal M}_{SS^*}|^2). \label{SDMself}
\ee
with 
\bea
|{\cal M}_{SS}|^2&=&2\left|2\Big(\lambda_S+\frac{18g^2_D m^2_S}{m^2_V}\Big)  +\frac{R^2_1 m^2_\chi}{4m^2_S-m_\chi^2+i\Gamma_\chi m_\chi }\right|^2, \nonumber \\
|{\cal M}_{SS^*}|^2&=&4\left|2\Big(\lambda_S-\frac{9g^2_D m_S^2}{m_V^2}\Big) -R^2_1 \right|^2. \nonumber
\eea
Here, we note that unitarity bounds on self-scattering are $|{\cal M}_{SS}|, |{\cal M}_{SS^*}|<8\pi$.

\begin{figure*}[!t]
  \begin{center}
   \begin{tabular}{cccccccc} 
   \includegraphics[height=0.43\textwidth]{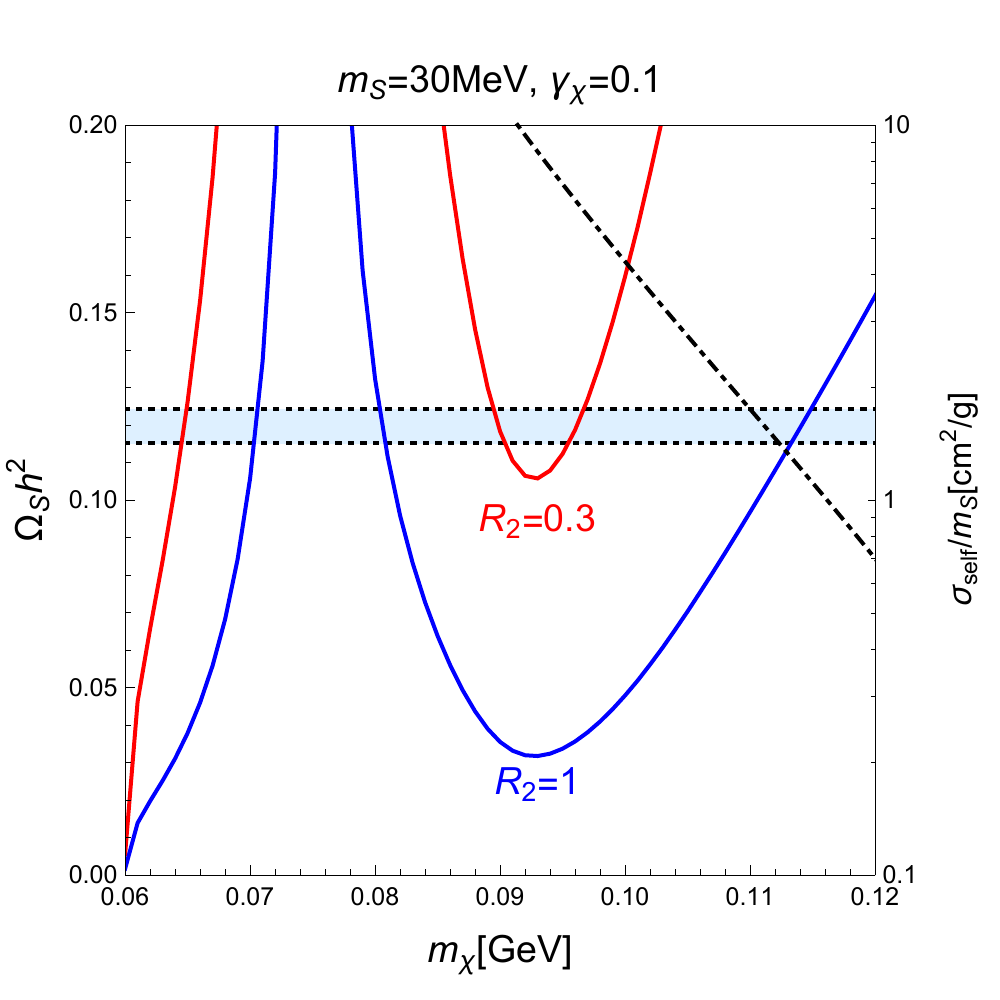}
  \includegraphics[height=0.43\textwidth]{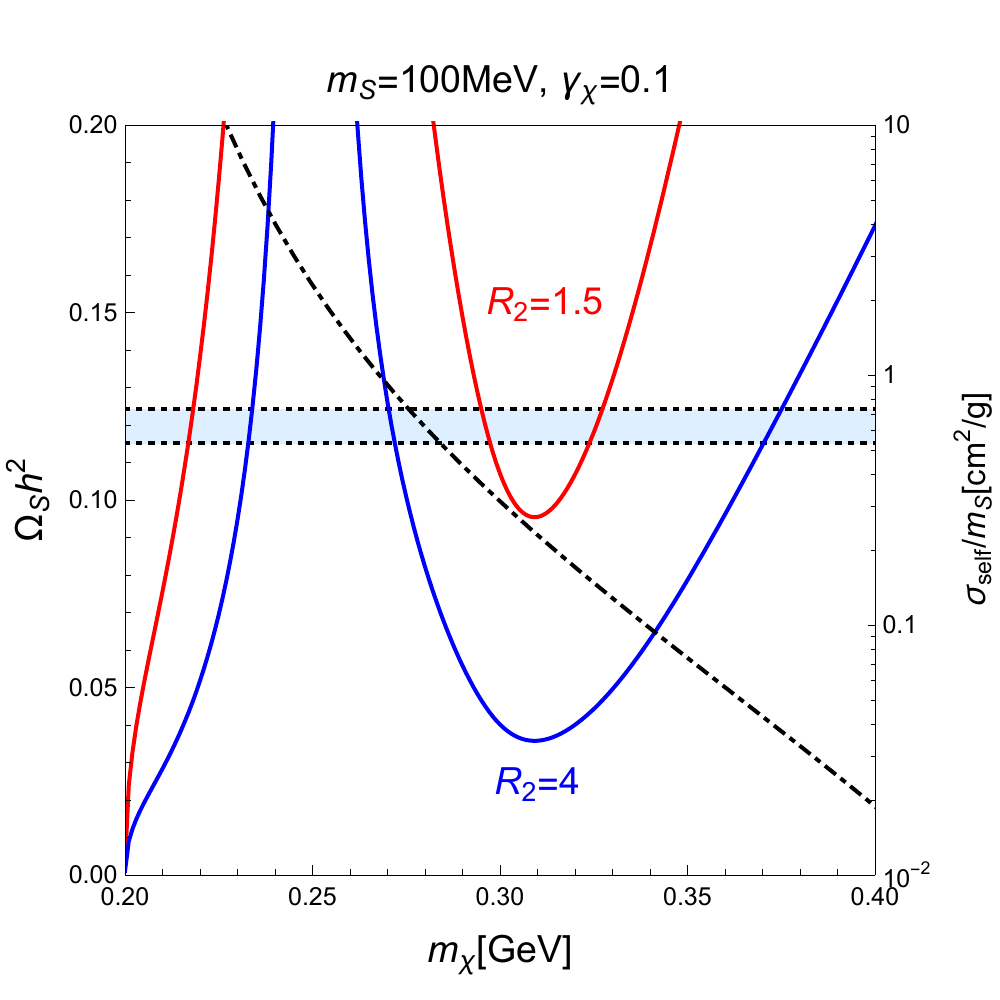} \\
   \includegraphics[height=0.43\textwidth]{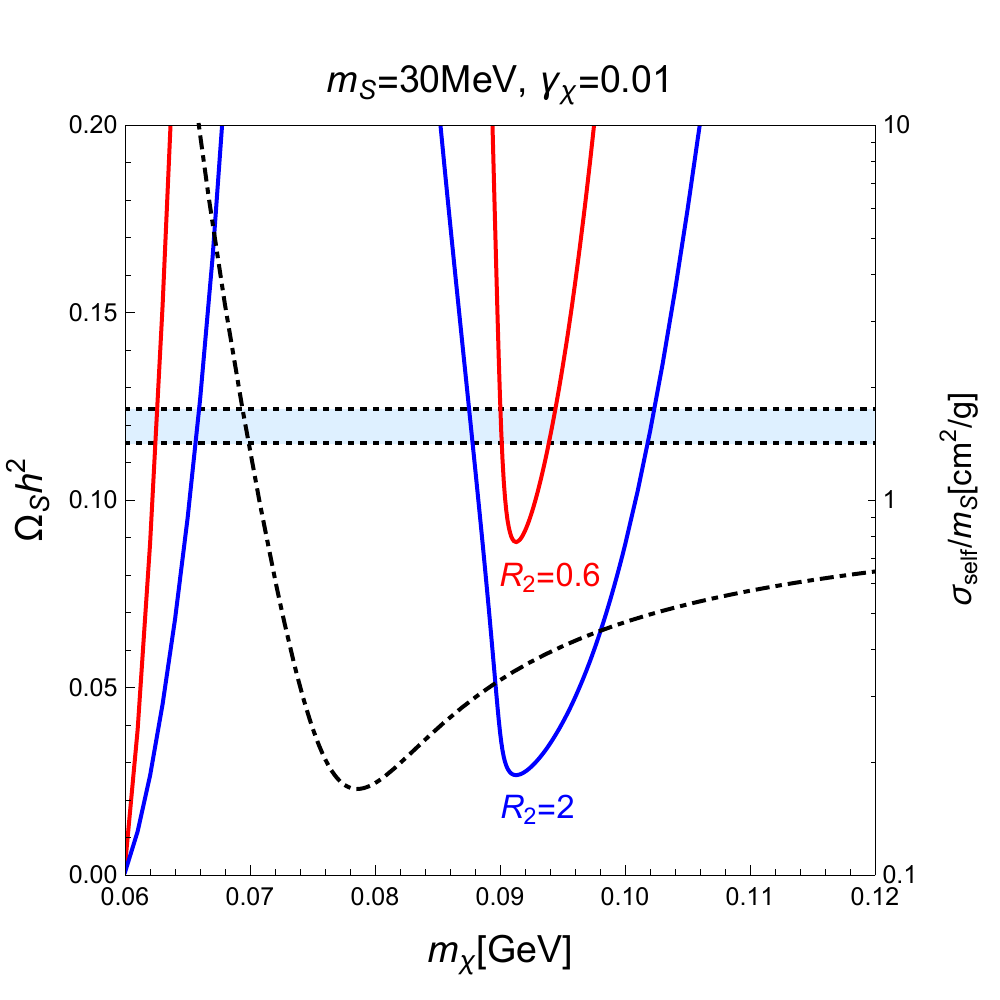}
   \includegraphics[height=0.43\textwidth]{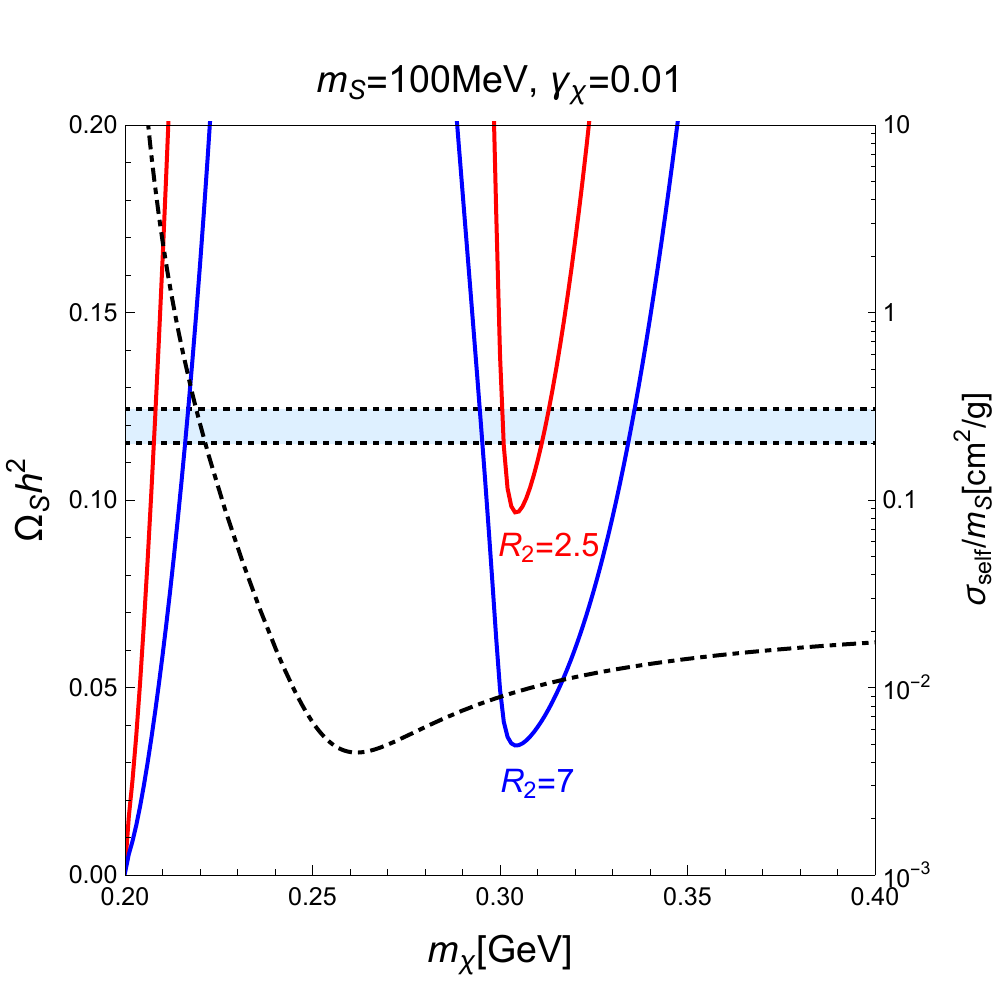}
  \end{tabular}
   \end{center}
  \caption{Relic density of $S$ SIMP as a function of the mediator mass $m_\chi$. We took $R_2$ of order one in solid lines. Planck $3\sigma$ band on the relic density is imposed in  horizontal light-blue region. For $m_V=500\,{\rm MeV}$,  $g_D=0.1$ and $\lambda_S=1$, self-scattering cross section ($\sigma_{\rm self}/m_S$) is shown in units of $1{\rm cm^2/g}$  in dot-dashed line. 
  }
  \label{temps}
\end{figure*}

Like the case with $m_S>2 m_\chi$, the annihilation cross section for $SSS\rightarrow S^*S^*$ is enhanced near the resonance with $m_\chi= 3m_S$.  Including a nonzero velocity of dark matter in the center of mass energy $s= 9m^2_S(1+v^2_{\rm rel}/4)$ in the propagators in eq.~(\ref{3to2}), the annihilation cross section for $SSS\rightarrow S^*S^*$ before thermal average has a temperature-dependent pole as follows,
\bea
(\sigma v^2)_{S} \equiv\frac{c_S}{m^5_S} \frac{\gamma^2_\chi}{(\epsilon_\chi-v^2_{\rm rel}/4)^2+\gamma^2_\chi}
\eea
where $\gamma_\chi\equiv m_\chi \Gamma_\chi/(9m^2_S)$, $\epsilon_\chi\equiv (m^2_\chi-9m^2_S)/(9m^2_S)$ parametrizes the off-set from the resonance pole, and $c_S$ is a constant parameter.  
Then, we can apply the results obtained for $\chi$ SIMP dark matter.

In Fig.~\ref{temps}, we show the relic density of the $S$ SIMP dark matter, depending on the SIMP mass and scalar coupling, $R_2$, of order one, in solid lines, and also depict the self-scattering cross section per SIMP mass, $\sigma_{\rm self}/m_S$, in dot-dashed lines.    
Similarly to the $\chi$ SIMP case, we find that  the relic density changes significantly, depending on the mass and width of the resonance.  We note that the correct relic density for $S$ SIMP can be obtained for a relatively small SIMP coupling $R_2$ than in the case of $\chi$ SIMP, because the width of the heavy scalar depends on SIMP mass and coupling differently.  The bounds on the self-scattering cross section constrains the parameter space, similarly to the case of $\chi$ SIMP. 

We remark on the other relations between singlet scalar masses. Namely, when $m_\chi<m_S< 2m_\chi$ or $m_\chi/2<m_S<m_\chi$, both $\chi$ and $S$ can be stable and become dark matter candidates. 
But, the heavier singlet scalar annihilates into the lighter one due to strong $2\rightarrow 2$ (semi-)annihilations, so the lighter singlet scalar becomes a dominant component of the observed relic density  \cite{workinprogress}. Nonetheless, we need to take into account extra $3\rightarrow 2$ annihilation channels: $\chi\chi\chi\rightarrow \chi S^*$, $\chi\chi\chi^*\rightarrow SS$ and $\chi\chi\chi^*\rightarrow \chi^* S^*$ for $m_\chi<m_S<2m_\chi$, and  $SSS\rightarrow \chi S$, $SS^*S^*\rightarrow \chi\chi$ and $SSS^*\rightarrow \chi S^*$ for $m_\chi/2<m_S<m_\chi$ \cite{workinprogress}.
For these mass relations, however, there is no resonant enhancement of the $3\rightarrow 2$ annihilation, unlike the cases with $2m_\chi< m_S$ or $2m_S<m_\chi$.

\section{Kinetic equilibrium via dark gauge boson}

SIMP dark matter could be problematic for structure formation \cite{structure}, unless the overheat coming from the $3\rightarrow 2$ annihilation process is equilibrated by the scattering with the SM thermal bath, namely, being in kinetic equilibrium.  When the dark gauge boson mixes with the SM hypercharge gauge boson by a gauge kinetic mixing term, ${\cal L}_{\rm kin}=-\frac{\varepsilon}{2\cos\theta_W} V_{\mu\nu} F^{\mu\nu}$ where $F_{\mu\nu}$ is the field strength for the SM hypercharge gauge boson, the elastic scattering cross section between $\chi(S)$ SIMP dark matter and a SM charged lepton for $\chi(S) l^\pm \rightarrow \chi(S) l^\pm$ is given by 
\be
\langle\sigma v\rangle_{\rm scatt, l^\pm}= \frac{24\pi q^2_{\chi(S)} \alpha_D \alpha\, \varepsilon^2 m^2_{\chi(S)}}{m^4_V}\,\Big(\frac{T}{m_\chi} \Big).
\ee
As a result, there appears a nonzero cross section for the $2\rightarrow 2$ annihilation of SIMP dark matter, $\chi\chi^*(SS^*)\rightarrow l^+ l^-$, as follows,
\be
\langle \sigma v\rangle_{\rm ann,l^+l^-}=\frac{32\pi q^2_{\chi(S)}\alpha_D \alpha\,\varepsilon^2 m^2_{\chi(S)}}{(4m^2_{\chi(S)}-m^2_V)^2} \, \Big(\frac{T}{m_\chi} \Big).
\ee
We note that the cross sections for kinetic scattering and $2\rightarrow 2$ annihilation for $S$ SIMP are nine times larger than those for $\chi$ SIMP, because the dark charges are $q_\chi=+1$ and $q_S=+3$. Then, the condition for kinetic equilibrium is $10^{-8}(m_V/m_{\chi(S)})^2\lesssim |q_{\chi(S)}\varepsilon|\lesssim 10^{-4}\sqrt{(m^2_V/m^2_{\chi(S)}-4)^2+\Gamma^2_V m^2_V/m^4_{\chi(S)}}$ for $\chi(S)$ SIMP dark matter. 
As  the annihilation cross section of SIMP dark matter is $p$-wave suppressed, there is no limit from current indirect detection experiments with cosmic rays \cite{z3dm}. On the other hand, the elastic scattering between SIMP dark matter and nucleon/electron could be constrained by direct detection experiments such as superconducting detectors \cite{superconductor,z3dm}. 

When $V$ decays invisibly into a pair of SIMP dark matter, the limits from $Z'$ searches with invisible decays are applicable as in the SIMP meson and $Z_3$ cases \cite{simp2a,z3dm}. 
However, there is a novel $V$ decay mode,  when $V$ decays into a pair of the heavy singlet scalars. 
For instance, for $\chi$ SIMP dark matter, the dark gauge boson decays  dominantly in cascade as $V\rightarrow SS^*\rightarrow \chi\chi\chi^*\chi^*$ for $m_V>2m_S\sim 6 m_\chi$. In this case, since the heavy scalar $S$ couples more strongly to $V$, the invisible decay width is larger than what we would expect from the direct decay, $V\rightarrow \chi\chi^*$.

\section{Conclusions}

We have proposed a model with discrete $Z_5$ gauge symmetry for SIMP scalar dark matter where the required $3\rightarrow 2$ annihilation cross section can be obtained without large couplings, due to the resonance of an additional scalar field.  We showed that when the width of the resonance gets smaller, there is a large parameter space of the SIMP masses and interactions in the perturbative regime,  satisfying the correct relic density and the bounds on the self-scattering cross section. Our model with two complex scalars shows the variety of the hidden dynamics. In particular, the invisible decay of the dark gauge boson can be boosted by the presence of the heavy singlet scalar resonance. 

\section*{Acknowledgments}

The work of HML is supported in part by Basic Science Research Program through the National Research Foundation of Korea (NRF) funded by the Ministry of Education, Science and Technology (2013R1A1A2007919). The work of SMC is supported by the Chung-Ang University Graduate Research Scholarship in 2016.

\end{document}